\documentclass[aps,prx,showpacs,floatfix,twocolumn,superscriptaddress,longbibliography]{revtex4-2}

\usepackage{float}
\usepackage{color}
\usepackage{bm}
\usepackage{hyperref}
\usepackage{todonotes}
\usepackage{verbatim}
\usepackage{soul}
\usepackage{glossaries}
\usepackage{sidecap}
\usepackage{hyperref}
\hypersetup{
    colorlinks=true,
    linkcolor=blue,
    filecolor=magenta,      
    urlcolor=red,
    citecolor=blue,
}

\def\Q{\ensuremath{\bm{Q}}}
\def\La438{La$_{4}$Ni$_{3}$O$_{8}$}
\def\Pr438{Pr$_{4}$Ni$_{3}$O$_{8}$}
\def\Ni112{\textit{R}NiO$_2$}
\def\LSCO{La$_{2-x}$Sr$_{x}$CuO$_{4}$}

\newacronym{BS}{BS}{bond-stretching}
\newacronym{RIXS}{RIXS}{resonant inelastic x-ray scattering}
\newacronym{XAS}{XAS}{X-ray absorption spectrum}
\newacronym{EELS}{EELS}{electron energy loss spectroscopy}
\newacronym{EPC}{EPC}{electron-phonon coupling}
\newacronym{CDW}{CDW}{charge density wave}
\newacronym{SDW}{SDW}{spin density wave}
\newacronym{FWHM}{FWHM}{full-width at half-maximum}
\newacronym{INS}{INS}{inelastic neutron scattering}
\newacronym{DFT}{DFT}{density functional theory}
\newacronym{GGA}{GGA}{generalized gradient approximation}
\newacronym{UHB}{UHB}{upper Hubbard band}
\newacronym{ZSA}{ZSA}{Zaanen-Sawatzky-Allen}
\newacronym{ZRS}{ZRS}{Zhang-Rice singlet}
\newacronym{ED}{ED}{exact diagonalization}
\newacronym{CEF}{CEF}{crystal electric field}
\newacronym{2D}{2D}{two-dimensional}
\newacronym{TM}{TM}{transition-metal}
\newacronym{DMFT}{DMFT}{dynamical mean field theory}

\begin{document}

\title{Role of Oxygen States in the Low Valence Nickelate La$_4$Ni$_3$O$_8$}

\author{Y. Shen}\email[]{yshen@bnl.gov}
\affiliation{Condensed Matter Physics and Materials Science Department, Brookhaven National Laboratory, Upton, New York 11973, USA}

\author{J. Sears}
\affiliation{Condensed Matter Physics and Materials Science Department, Brookhaven National Laboratory, Upton, New York 11973, USA}

\author{G. Fabbris}
\affiliation{Advanced Photon Source, Argonne National Laboratory, Lemont, Illinois 60439, USA}

\author{J. Li}
\author{J. Pelliciari}
\author{I. Jarrige}
\affiliation{National Synchrotron Light Source II, Brookhaven National Laboratory, Upton, New York 11973, USA}

\author{Xi He}
\author{I. Bo\v{z}ovi\'{c}}
\affiliation{Condensed Matter Physics and Materials Science Department, Brookhaven National Laboratory, Upton, New York 11973, USA}
\affiliation{Department of Chemistry, Yale University, New Haven, Connecticut 06520, USA}

\author{M. Mitrano}
\affiliation{Department of Physics, Harvard University, Cambridge, Massachusetts 02138, USA}

\author{Junjie Zhang}
\affiliation{Materials Science Division, Argonne National Laboratory, Lemont, Illinois 60439, USA}
\affiliation{Institute of Crystal Materials, Shandong University, Jinan, Shandong 250100, China}

\author{J. F. Mitchell}
\affiliation{Materials Science Division, Argonne National Laboratory, Lemont, Illinois 60439, USA}

\author{A. S. Botana}
\affiliation{Department of Physics, Arizona State University, Tempe, Arizona 85287, USA}

\author{V. Bisogni}
\affiliation{National Synchrotron Light Source II, Brookhaven National Laboratory, Upton, New York 11973, USA}

\author{M. R. Norman}
\affiliation{Materials Science Division, Argonne National Laboratory, Lemont, Illinois 60439, USA}

\author{S. Johnston}
\affiliation{Department of Physics and Astronomy, The University of Tennessee, Knoxville, Tennessee 37966, USA}
\affiliation{Institute of Advanced Materials and Manufacturing, The University of Tennessee, Knoxville, Tennessee 37996, USA}

\author{M. P. M. Dean}\email[]{mdean@bnl.gov}
\affiliation{Condensed Matter Physics and Materials Science Department, Brookhaven National Laboratory, Upton, New York 11973, USA}

\date{\today}

\begin{abstract}
The discovery of superconductivity in square-planar low valence nickelates has ignited a vigorous debate regarding their essential electronic properties: Do these materials have appreciable oxygen charge-transfer character akin to the cuprates, or are they in a distinct Mott-Hubbard regime where oxygen plays a minimal role? Here, we resolve this question using O $K$-edge \gls*{RIXS} measurements of the low valence nickelate \La438{} and a prototypical cuprate \LSCO{} ($x=0.35$). As expected, the cuprate lies deep in the charge-transfer regime of the \gls*{ZSA} scheme. The nickelate, however, is not well described by either limit of the \gls*{ZSA} scheme and is found to be of mixed charge-transfer/Mott-Hubbard character with the Coulomb repulsion $U$ of similar size to the charge-transfer energy $\Delta$. Nevertheless, the transition-metal-oxygen hopping is larger in \La438{} than in \LSCO{}, leading to a significant superexchange interaction and an appreciable hole occupation of the ligand O orbitals in \La438{} despite its larger $\Delta$. Our results clarify the essential characteristics of low valence nickelates and put strong constraints on theoretical interpretations of superconductivity in these materials.
\end{abstract}

\maketitle

\section{Introduction}

\begin{figure*}
\includegraphics{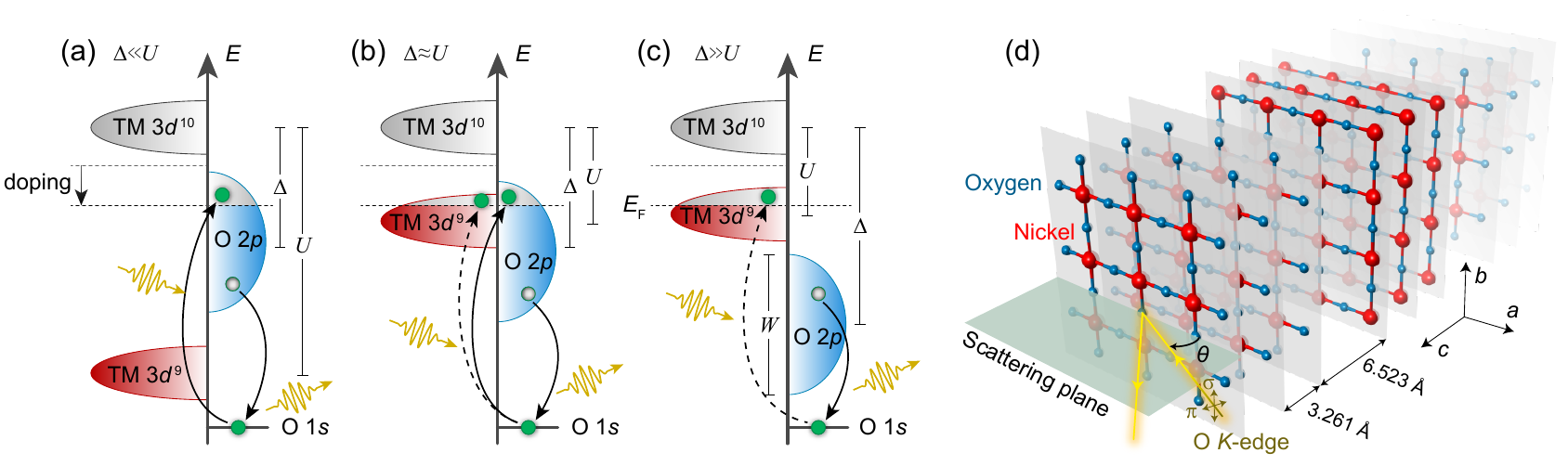}
\caption{Schematic of the O $K$-edge RIXS process. (a)--(c) RIXS processes for different values of the Hubbard $U$ and charge-transfer energy $\Delta$. These are defined as the energy cost of specific atomic transitions in the material: for a $d^9$ ground state, $U$ being a $d_i^9d_j^9\rightarrow d^{8}_id_j^{10}$ and $\Delta$ being a $d_i^9\rightarrow d^{10}_i\underline{L}$ transition, where $i$ and $j$ label \gls*{TM} sites and $\underline{L}$ denotes an oxygen ligand hole. For a charge-transfer insulator ($\Delta \ll U$), the doped holes are mostly in the oxygen $2p$ orbitals, while in a Mott-Hubbard insulator ($\Delta \gg U$), the doped holes mainly occupy the \gls*{TM} $3d^n$ state. In the mixed charge-transfer/Mott-Hubbard regime ($\Delta \sim U$), the doped holes are spread among both the \gls*{TM} and oxygen sites. Arrows show examples of x-ray transition pathways, which, because of x-ray dipole selection rules, can involve either O states or \gls*{TM}-O hybridized states, making this process ideal to distinguish situations (a)--(c). Note that $W$ is the band width for oxygen $2p$ orbitals, and $E_{\textrm{F}}$ is the Fermi energy (marked by the horizontal dashed lines). (d) Experimental setup for RIXS measurements at the O $K$-edge and crystal structure of \La438{}. Only the nickel and oxygen ions in the trilayers are shown.}
\label{fig:schematic}
\end{figure*}

Creating analogs of the cuprate high-temperature superconductors has been a target of materials research for decades \cite{Anisimov1999electronic, Norman2016materials, Adler2018correlated}. It is widely believed that this requires a formal $d^{9-\delta}$ electron count on the \gls*{TM} site (as in hole-doped cuprates), strong electronic correlations, and a substantial \gls*{TM}-O hybridization. This situation is illustrated in Fig.~\ref{fig:schematic}(a), where the Cu-O charge-transfer energy $\Delta$ in cuprates is much smaller than the on-site $3d$ Coulomb repulsion $U$. Square-planar nickelate materials of the form $R_{n+1}$Ni$_{n}$O$_{2n+2}$ (\textit{R} stands for a rare earth and $n$ is the number of neighboring NiO$_2$ layers) realize a $d^{9-\delta}$ valence and were identified as promising potential cuprate analogs \cite{Anisimov1999electronic, Poltavets2010bulk}. This strategy bore fruit when infinite-layer $n=\infty$ \textit{R}$_{1-x}$Sr$_x$NiO$_2$ materials, and more recently, $n=5$ Nd$_6$Ni$_5$O$_{12}$, were shown to superconduct, generating intense scientific interest \cite{Li2019superconductivity, Osada2020superconducting, Zeng2021superconductivity, Pan2021super}.

Even before this discovery, however, it was noted that Ni has a smaller nuclear charge than Cu, which would tend to increase $\Delta$ \cite{Lee2004infinite}. If $\Delta$ is substantially larger, this could potentially realize a Mott-Hubbard situation, as illustrated in Fig.~\ref{fig:schematic}(c), in which O states play a minimal role in the low-energy physics. This expectation motivated extensive interest in determining the Mott-Hubbard versus charge-transfer nature of square-planar nickelates, with some researchers categorizing them as Mott-Hubbard systems, and others emphasizing the nickelates as cuprate analogs \cite{Hu2019two,Nomura2019formation, Sakakibara2019model, Sawatzky2019superconductivity, Adhikary2020orbital, Bixia2020synthesis, Botana2020similarities, choi2020fluctuation, Hepting2020electronic, Jiang2020critical, Lechermann2020late, Lechermann2020multiorbital, Liu2020electronic, Petocchi2020normal,
Karp2020many, Karp2020comparative, Kapeghian2020electronic, Nomura2020magnetic, Wang2020hunds, Werner2020nickelate, Wu2020robust, Zhang2020effective, Zhang2020self, Zhang2020type, Zeng2020phase, Been2021electronic, Goodge2021doping, Lang2021strongly, kang2021optical, Plienbumrung2021interplay, Wan2021exchange, Higashi2021core}. Resolving this controversy is a crucial step for understanding these materials. The degree of \gls*{TM}-O hybridization is a major determining factor of the relative Ni versus O character of the charge carriers and the magnetic superexchange. This issue also constrains which minimal effective models are appropriate for these materials and is therefore central to understanding nickelate superconductivity. Part of the challenge of this question is that $\Delta$, $U$, the hopping integrals $t_{pd}$ and $t_{pp}$, and the crystal fields all act together to determine the role of oxygen and the location of the material within the \acrfull*{ZSA} scheme \cite{ZSA1985, ZSA1987}. Progress in this area has also been hindered by varying interpretations based on different experimental approaches \cite{Lu2021magnetic, Rossi2020orbital, Fu2019core, Hepting2020electronic, Gu2020single, Goodge2021doping, Chen2021electronic, Higashi2021core} and the fact that many-body electronic structure results are sensitive to the method used \cite{Karp2021dependence}. This controversy has been further amplified by the recent discovery of relatively large superexchange interactions, about half that of cuprates, in both the $n=3$ and $n=\infty$ materials \cite{Lin2021strong, Lu2021magnetic}. Resolving these issues requires a spectroscopic probe that can specifically target oxygen states and their energies and an interpretative approach that treats $\Delta$, $U$, $t_{pd}$, and $t_{pp}$ on the same footing. 

Here, we use O $K$-edge \gls*{RIXS} to determine the Mott-Hubbard versus charge-transfer electronic characteristics of the $n=3$ nickelate \La438{}, which can be prepared as a bulk single crystal without the need for chemical doping, and the oxygen-containing substrates and capping layers typically required for the  stability of the infinite-layer nickelate superconductors. We benchmark our results for this material against a prototypical cuprate \LSCO{} ($x=0.35$), which has a similar effective doping. The nickelate is found to be of mixed charge-transfer/Mott-Hubbard electronic character and not well described by either the Zhang-Rice singlet states found in \LSCO{} nor a Ni-dominated Mott-Hubbard scenario. We confirm this interpretation using \gls*{ED} calculations and further find that the \gls*{TM}-oxygen hopping is enhanced in \La438{} relative to \LSCO{}, with a $\Delta$ about 2~eV larger. These parameters lead to a sizable superexchange interaction and an appreciable O character of the doped holes. Our results clarify the essential electronic characteristics of nickelates and imply that both O and Ni are necessary ingredients of minimal effective models for these materials. 

\section{Methods}


High-energy-resolution \gls*{RIXS} measurements were performed at the SIX beamline at the NSLS-II with an energy resolution of around 22~meV. All data shown were collected at 40~K. We fixed the spectrometer at a horizontal scattering angle $2\Theta=150^\circ$ and used either $\pi$ or $\sigma$ polarized x-rays as shown in Fig.~\ref{fig:schematic}(d). Thus, we put the crystalline ($H$, 0, 0) and (0, 0, $L$) directions in the scattering plane and changed \Q{} by rotating the sample around the vertical (0, 1, 0) axis. Note that $\theta$ is defined as the angle between the incident beam and sample surface, which is perpendicular to the sample $c$ axis.


Our chosen low-valence nickelate sample is \La438{} ~\cite{Zhang2016stacked, Zhang2017large, Zhang2019stripe, Bernal2019charge, Lin2021strong}. This material has a nominal Ni $d^{8\frac{2}{3}}$ valence and can be considered as $1/3$ self-doped with holes. Here, ``doped holes" refers to this $\delta=1/3$ additional holes with respect to $d^9$, and ``holes" refers to the total holes with respect to the closed Ni $d$ shell. With no alkaline earth substitution on the rare-earth site, the \La438{} system presents a well-defined hole doping and avoids sample inhomogeneity challenges that have hindered studies of \textit{R}$_{1-x}$Sr$_x$NiO$_2$ \cite{Goodge2021doping, Osada2021nickelates, Zeng2021observation, Si2020topotactic, Malyi2021bulk, Higashi2021core}.

Single crystal samples were reduced from the as-grown Ruddlesden-Popper phase La$_4$Ni$_3$O$_{10}$ using H$_2$/Ar gas \cite{Zhang2016stacked, Zhang2017large, Zhang2019stripe} (see Appendix~\ref{synthesis} for details). \La438{} has a tetragonal structure with space group $I4/mmm$ and lattice constants of $a=b=3.97$ \AA, $c=26.1$ \AA. Its structure features trilayers of square-lattice NiO$_2$ planes, which are separated by La$_2$O$_2$ fluorite blocks [see Fig.~\ref{fig:schematic}(d)]. All members of the $R_{n+1}$Ni$_{n}$O$_{2n+2}$ family share very similar Ni-O bonding and would therefore be expected to have similar local correlated physics, provided they are compared at the same effective doping. This expectation is borne out by calculations \cite{Karp2020comparative, LaBollita2021electronic} and the similar magnetic exchange \cite{Lin2021strong, Lu2021magnetic} and superconducting transition temperature in different nickelates \cite{Li2019superconductivity, Osada2020superconducting, Zeng2021superconductivity, Pan2021super}. We compare \La438{} to \LSCO{} ($x=0.35$) samples prepared via molecular beam epitaxy with a similar effective doping as the nickelate.

\begin{figure}[b]
\includegraphics{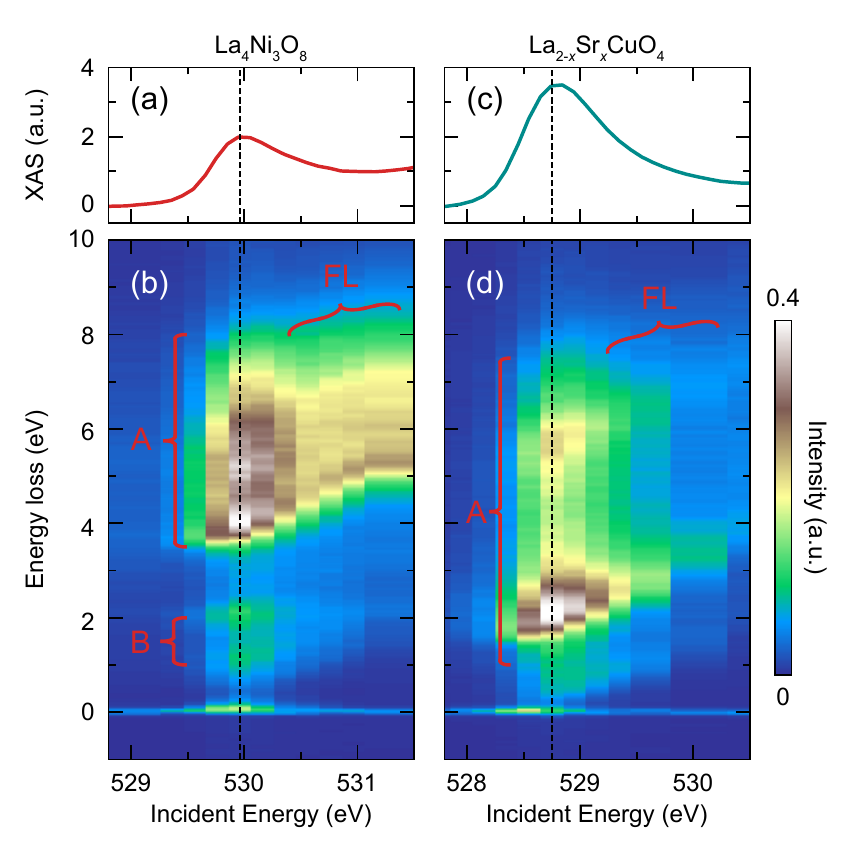}
\caption{Comparison of \gls*{RIXS} spectra in \La438{} and \LSCO{}. \gls*{XAS} plots and \gls*{RIXS} intensity maps across the pre-peak energies for (a,b) \La438{} and (c,d) \LSCO{}. The data were collected with $\theta=15^\circ$ and $\sigma$ polarization. Different contributions can be identified, including the fluorescence (FL) and resonant signals (labeled as features A and B). The vertical dashed lines indicate the pre-peak resonant energy. \gls*{XAS} data are shown in arbitrary units (a.u.) normalized to 0 and 1 at energies 1~eV below and 1~eV above the pre-peak \cite{Zhang2017large, Lin2021strong}.}
\label{fig:Emap}
\end{figure}

\begin{figure}[b]
\includegraphics{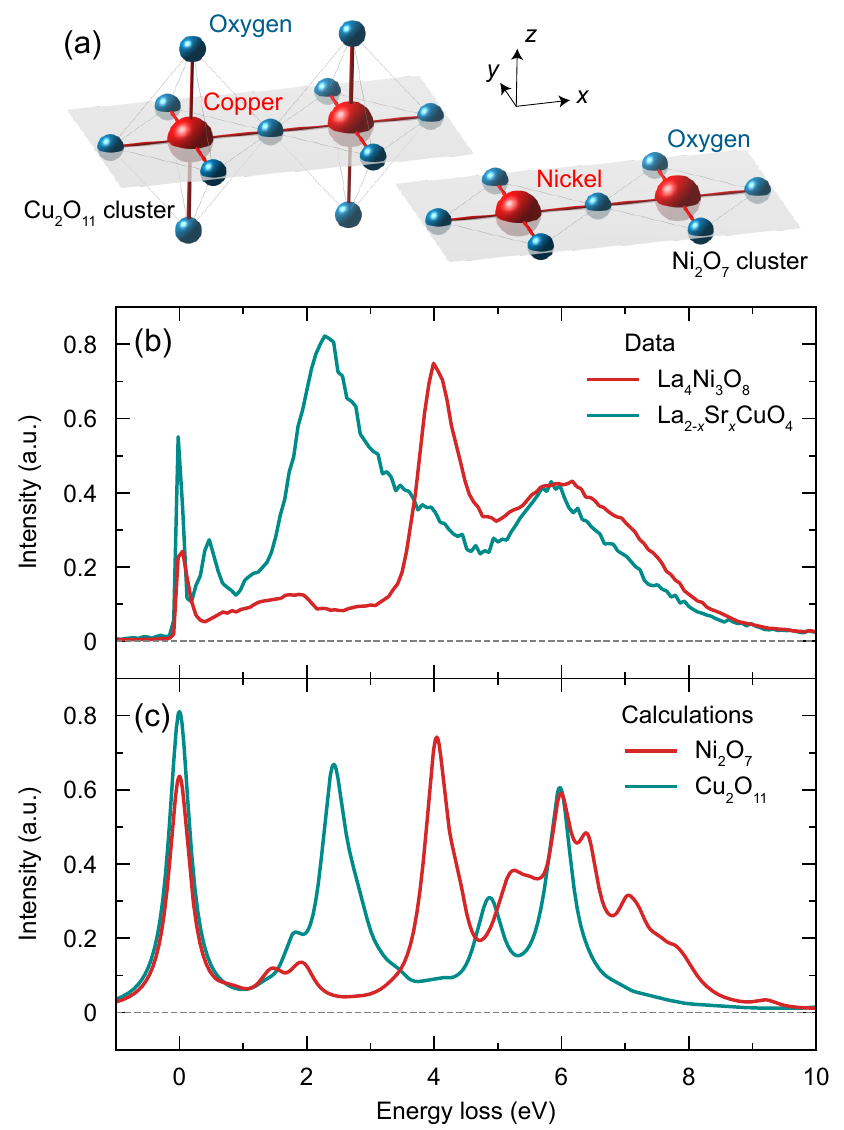}
\caption{Cluster exact diagonalization results. (a) Sketch of the clusters used in each case. (b) \gls*{RIXS} curves measured at the O pre-peak resonance with $\theta=60^\circ$ and $\sigma$ polarization. The feature at 0.5~eV in the \LSCO{} sample is known to come from a plasmon, so is not expected to appear in our model calculations \cite{Nag2020detection}. (c) Calculated spectra using the parameters listed in Table~\ref{table:params} with the same experimental geometries.}
\label{fig:EDcalc}
\end{figure}

\section{Results and interpretation}


We start by measuring x-ray absorption spectra of our nickelate and cuprate samples in the O $K$ pre-edge region, as shown in Figs.~\ref{fig:Emap}(a) and \ref{fig:Emap}(c). The spectra have a pre-peak feature, which comes from holes occupying \gls*{TM}-O hybridized states \cite{Chen1991electronic, Chen1992out, Zhang2017large, Lin2021strong}. As measured previously, this pre-peak is weaker in \La438{} compared to \LSCO{}, but this in itself is insufficient to quantify the electronic nature of these materials in detail \cite{Zhang2017large, Lin2021strong}. 

Figures~\ref{fig:Emap}(b) and \ref{fig:Emap}(d) show the \gls*{RIXS} energy maps at $\theta=15^\circ$ with $\sigma$ polarization. The maps consist of strong resonant features that appear at the O $K$ pre-peak and diagonal structures of weaker, more diffuse intensity above the resonance. Features that appear as diagonal lines in these maps are assigned to x-ray fluorescence (labeled FL), which occur at constant final energy and arise from a continuum of particle-hole excitations. Apart from the fluorescence line, Raman-like peaks are observed at the pre-peak incident energy in both samples, which correspond to local excitations. For \La438{}, the resonant signals can be divided into a strong manifold of excitations between 4 and 8~eV (feature A) and a low-energy weaker structure between 1 and 2~eV (feature B) [see Figs.~\ref{fig:Emap}(b) and \ref{fig:EDcalc}(b)]. For \LSCO{}, one manifold of excitations, which we call feature A, covers 1--8~eV, making feature B hard to distinguish [see Figs.~\ref{fig:Emap}(d) and \ref{fig:EDcalc}(b)]. 

\begin{table*}[hbt]
\caption{Representative parameters used for the ED calculations. Full details are provided in Appendix~\ref{EDcalcs}. Note that $U$ is the on-site intra-orbital Coulomb interaction for the $3d$ orbitals, and $\Delta$ measures the charge-transfer energy. These parameters have an estimated error bar of about 1~eV \cite{supp}. Here, $t_{p_{\sigma}d_{x^2-y^2}}$ is the hopping integral between planar $p_{\sigma}$ and $d_{x^2-y^2}$ orbitals while $t_{p_{\sigma}p_{\sigma}}$ is that among the $p_{\sigma}$ orbitals. The column labeled $J_{\textrm{calc}}$ is the singlet-triplet energy splitting of the undoped ($d^9$) cluster (two holes in total), and $J_{\textrm{exp}}$ is the superexchange interaction from experiment. These parameters are given in eV. The column labeled \gls*{TM} holes is the hole occupation of the \gls*{TM} orbitals per \gls*{TM} site of the doped 2-\gls*{TM} site clusters (which have three holes in total). The column labeled O holes is the hole occupation of the oxygen orbitals per \gls*{TM} site. }
\centering
\begin{ruledtabular}
\begin{tabular}{ccccccccc}
Cluster & $U$ & $\Delta$ & $t_{p_{\sigma}d_{x^2-y^2}}$ & $t_{p_{\sigma}p_{\sigma}}$ & $J_{\textrm{calc}}$ & $J_{\textrm{exp}}$ & TM holes & O holes  \\ 
\hline
Ni$_2$O$_7$ & 6.5 & 5.6 & 1.36 & 0.375 & 0.084 & 0.069 \cite{Lin2021strong} & 1.05 & 0.45  \\
Cu$_2$O$_{11}$  & 9 & 3.4 & 1.17 & 0.625 & 0.151 & 0.143 \cite{Headings2010LCOINS} & 0.76 & 0.74 \\
\end{tabular}
\end{ruledtabular}
\label{table:params}
\end{table*}

In the photon emission process of O $K$-edge \gls*{RIXS}, electrons from either the \gls*{TM} $3d$ or oxygen $2p$ orbitals can potentially be deexcited to fill the core hole; however, the former channel is expected to be weaker since it is only possible through the \gls*{TM}-O hybridization. In light of this, we might expect feature A to come from charge-transfer excitations in the O $2p$ states and feature B to come from the weaker inter-orbital $dd$ excitations of the \gls*{TM} $3d$ orbitals. This is further verified by the Ni $L$-edge \gls*{RIXS} data we have taken, which show a strong peak in the 1--2~eV energy window.  The fact that feature A for \La438{} is higher in energy than that in \LSCO{} indicates a larger $\Delta$ by about 2~eV. At the same time, the presence of feature B from $dd$ excitations suggests that there is still appreciable Ni-O hybridization, which would not occur if $\Delta \gg U$. It is also clear that feature A is narrower in the energy loss axis in \La438{}, suggesting a smaller oxygen band width, again consistent with the larger $\Delta$. Considering these factors together allows us to conclude that \La438{} is of mixed charge-transfer/Mott-Hubbard character. 

To confirm and quantify our conclusion of mixed charge-transfer/Mott-Hubbard character, we make use of cluster \gls*{ED} calculations using the EDRIXS software \cite{Wang2019EDRIXS, EDRIXS} to reveal the electronic structure and oxygen involvement in \La438{} and \LSCO{}. The calculations are performed on Ni$_2$O$_7$ and Cu$_2$O$_{11}$ clusters, respectively, with open boundary conditions and three holes in total (i.e., 1.5 holes per \gls*{TM} site or $x=0.5$), as shown in Fig.~\ref{fig:EDcalc}(a). The orbital basis in each case includes all five \gls*{TM} $3d$ orbitals and all three O $2p$ orbitals on their respective atoms, and parameters are defined in hole notation such that $\Delta$ denotes the energetic splitting between the O and \gls*{TM} states (see Appendixes \ref{notation} and \ref{EDcalcs}). More specifically, $\Delta$ is defined as the energy cost for a local $d_i^9\rightarrow d^{10}_i\underline{L}$ transition, and $U$ reflects the energy needed for a $d_i^9d_j^9\rightarrow d^{8}_id_j^{10}$ transition\cite{ZSA1985, ZSA1987}. Although in undoped $d^9$ infinite-layer nickelates, the rare-earth orbitals act as effective dopants, \gls*{DMFT} calculations suggest that their role is minimal in heavily doped nickelates, where these states are unoccupied and only weakly hybridized with the planar Ni and O orbitals \cite{Karp2020comparative}. Omitting rare-earth orbitals allows us to maintain a tractable basis size while still achieving a satisfactory description of the data. Starting from the atomic limit, we explicitly include Coulomb interactions and nearest-neighbor inter-atomic hopping, and simulate the \gls*{XAS} and \gls*{RIXS} spectra with experimental conditions fully accounted for (see Appendix~\ref{EDcalcs} for more details). As explained in detail in the Supplemental Material \cite{supp}, the model parameters were then adjusted to match the experimental spectra, and they are listed in Table~\ref{table:params}. The strong core-hole potential in \gls*{RIXS} means that the low-energy excitations seen at resonance tend to have rather local character and minimal dispersion, which supports and motivates the widespread cluster-based interpretation of \gls*{RIXS} measurements \cite{Okada2003theory, Okada2002copper, Okada2004effects, Monney2013determin, Monney2016probing, Johnston2016electron, Plienbumrung2021character, Chen2013doping}.

Our approach fully incorporates many-body and multi-orbital effects, avoiding issues with \textit{ad hoc} parameters or double counting \cite{Dang2014covalency, Haule2015exact}, and it allows us to directly compare theory and experiment by evaluating the Kramers-Heisenberg equation for the \gls*{RIXS} cross section. In this way, we can extract $\Delta$ and $U$ in the original sense defined within the \gls*{ZSA} scheme as outlined in Fig.~\ref{fig:schematic}. Other means of extracting these parameters, such as those relying on density functional theory, can potentially be affected by double counting of the Coulomb interactions \cite{Dang2014covalency, Haule2015exact, supp}. These advantages come at the cost of having to work with a small cluster, which means that the effective doping level ($x=0.5$) is only approximately that of the real materials ($x\approx1/3$). Potential minor discrepancies are expected, arising from the small cluster size and omitted interactions with other atoms such as the rare-earth ions and adjacent layers. In addition, the cluster does not capture the FL feature since this requires a continuum of states \cite{Hariki2018continuum, Higashi2021core}. These factors lead to an error bar of about 1~eV on the extracted parameters, but within this error, the extracted parameters are robust, as has been demonstrated in several prior small cluster calculations of related oxide materials including cuprates and infinite-layer nickelates \cite{Okada2003theory, Okada2002copper, Okada2004effects, Monney2013determin, Monney2016probing, Johnston2016electron, Plienbumrung2021character, Chen2013doping}. 

The calculated \gls*{RIXS} spectra are shown in Fig.~\ref{fig:EDcalc}(c) and capture the onset and bandwidth of the spectral features nicely, including feature A, and the fact that feature B is difficult to observe in \LSCO{}. The observed level of agreement is comparable or surpasses the typical level of agreement of theoretical interpretations of O $K$-edge \gls*{RIXS} spectra \cite{Chen2013doping, Monney2013determin, Monney2016probing, Johnston2016electron, Jiang2020critical, Plienbumrung2021character}.

\begin{figure*}[bt]
\includegraphics{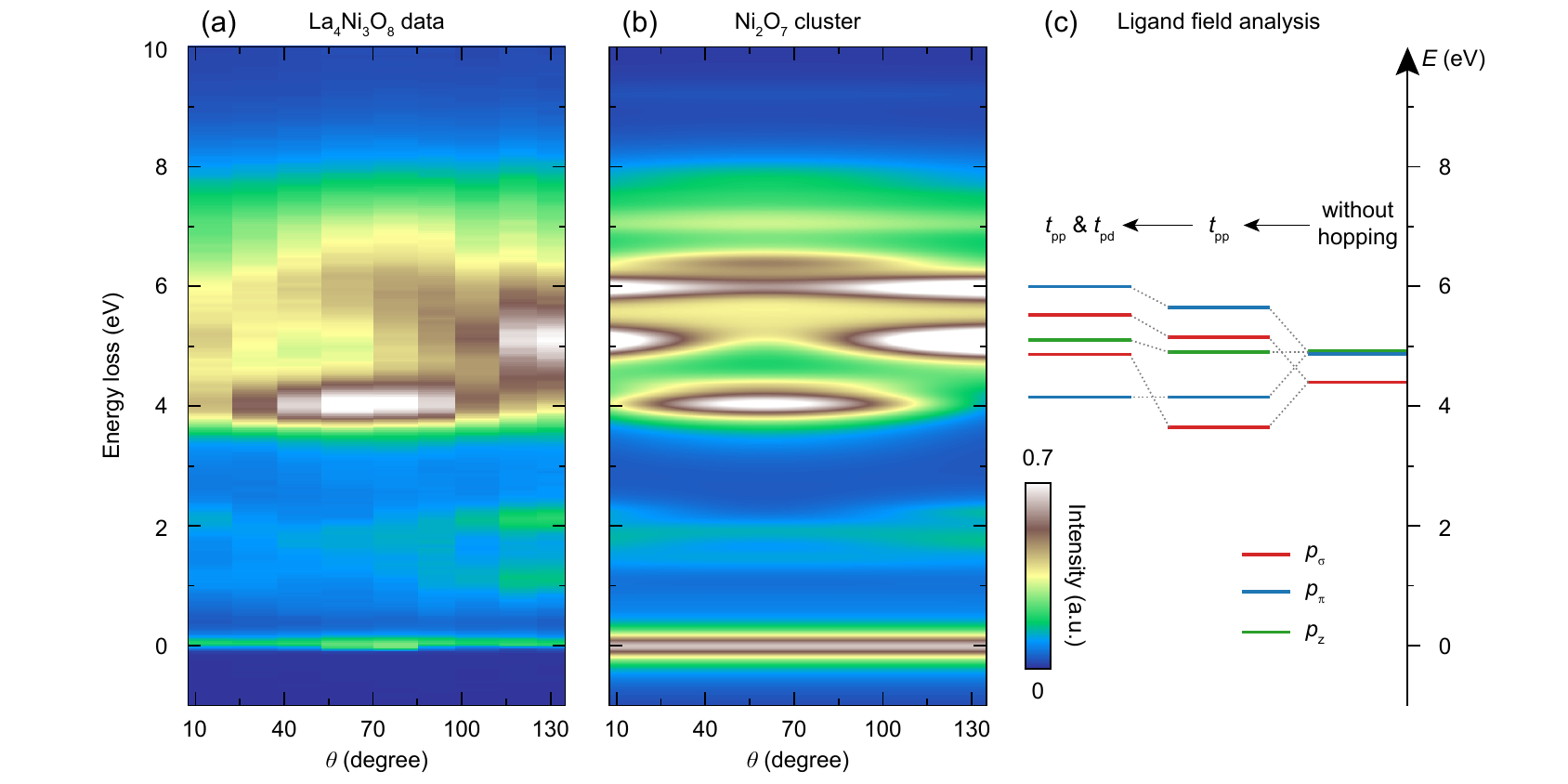}
\caption{Angle dependence of La$_{4}$Ni$_{3}$O$_{8}$ \gls*{RIXS} intensity at the pre-edge with $\sigma$ polarization. (a) \gls*{RIXS} map data after absorption correction. (b) Simulation of a Ni$_2$O$_7$ cluster. (c) Orbital energy-level diagram of the multiplet ligand field analysis using the parameters listed in Table \ref{table:params}. A full description of these states can be found in  Appendix~\ref{ligand_analysis}. Since this breakdown of the different orbitals is based on one-hole basis functions, the $y$ energy axis has been shifted to account for the doping.}
\label{fig:THdep}
\end{figure*}

Table~\ref{table:params} summarizes the parameters extracted from our theoretical analysis. As expected, these values confirm our empirical conclusions regarding the charge-transfer versus Mott-Hubbard characteristics of \La438{} and \LSCO{}. We find that $\Delta$ is much smaller than $U$ in \LSCO{}, confirming its charge-transfer nature, consistent with the known character of cuprates \cite{Anisimov1999electronic, Norman2016materials, Adler2018correlated}. Our $\Delta\sim3.4$~eV determination is very similar to the literature values from prior \gls*{ED} studies and within 1~eV (slightly larger) of those in first-principles work \cite{Weber2012Scaling, Chen2013doping}. This is consistent with the estimated 1~eV error bar of our parameter extraction and indicates that the finite-size cluster effect will only change a parameter by less than this error bar. In contrast, \La438{} has a significantly larger $\Delta\sim5.6$ eV, which is comparable to $U\sim6.5$~eV. This value places \La438{} in a mixed charge-transfer/Mott-Hubbard regime. It should be noted that different combinations of \gls*{ED} parameters can produce similar \gls*{RIXS} spectra for \La438{}, but all of them give a $\Delta\approx 6$~eV within a 1~eV error bar and a $U$ of a similar size to $\Delta$ \cite{supp}. As a consequence, the nickelate has enhanced Ni hole character compared to the cuprate, but it retains appreciable O hole character and therefore is not in the pure Mott-like regime in which the holes would have overwhelming \gls*{TM} $d$ character. Our fits to the data also find that the $p-d$ hopping is enhanced in \La438{}, which we attribute to the more extended nickel $d$ orbitals due to the smaller nuclear charge \footnote{Interestingly, this enhancement is not seen in a DFT Wannier analysis of the infinite-layer nickelate \cite{Botana2020similarities} and \La438{} \cite{Nica2020theoretical}, where a value similar to the cuprates is found.}. Coupled with the fact that $\Delta$ is only about 2~eV larger than in the cuprates, this means that a significant portion of the doped holes still reside on the oxygen sites (Table~\ref{table:params}). At the same time, \La438{} has a larger fraction of $|d^8\rangle$ states in its ground-state wave function compared to \LSCO{} because of its larger $\Delta$ and lower $U\approx \Delta$ (see Appendix~\ref{wavefunction}), as also found in \gls*{DMFT} studies \cite{Lechermann2020multiorbital, Karp2020comparative, Wang2020hunds}. Overall, the \gls*{TM}-O hybridization, which can be quantified by the parameter $t_{pd}^2/\Delta$, is slightly reduced in \La438{}: 0.33 for \La438{} compared to 0.4 for \LSCO{}. 

Examining the wave functions in Appendix~\ref{wavefunction}, we find that feature A in \La438{} is largely of a charge-transfer type where a significant fraction of a hole (about 0.43) is excited from the double-hole-occupied Ni ion to the ligand O orbitals (i.e., $|d^8\rangle \rightarrow |d^{9}\underbar{L}\rangle$). In \LSCO{}, the excitation is more mixed; it involves both a small amount of charge transfer (about 0.14 holes) from the Cu to ligand O and a rearrangement of the ligand hole (i.e., $|d^9\underbar{L}\rangle \rightarrow |d^9\underbar{L}^\prime\rangle$). Feature B, on the other hand, corresponds to $dd$ excitations, where the ground-state occupations of the \gls*{TM} $3d$ orbitals are rearranged during the scattering process. They are also present but weaker in the \LSCO{} calculations and obscured by feature A's  overlapping contribution. Detailed projections of the wave functions onto different states are provided in Appendix~\ref{wavefunction}.

To further verify our model for \La438{}, we show that our \gls*{ED} calculations also capture the angular dependence of the spectra. In this case, the evolution of the RIXS intensity with $\theta$ can largely be accounted for by the matrix elements of the \gls*{RIXS} cross section, which select different O orbitals. They can be easily calculated in the dipole approximation and the results are reasonably consistent with the \La438{} data (Fig.~\ref{fig:THdep}). The different levels can be conceptualized using a multiplet ligand field model \cite{Eskes1990cluster, Haverkort2012multiplet}, which is a simplified version of the cluster model we use for our full analysis, as explained in Appendix~\ref{ligand_analysis}. In this approach, the O $2p$ orbitals hybridize with each other, forming bonding, anti-bonding and non-bonding ligand orbitals through in-plane $p-p$ hopping. Some of these have the same symmetry as the \gls*{TM} $3d$ orbitals and thus hybridize with them via the relevant $p-d$ hopping. Using the parameters determined from our \gls*{ED} calculations, we produce a simplified scheme of the ligand orbital energy levels in Fig.~\ref{fig:THdep}(c). Although the ligand orbital splitting is slightly underestimated because of omitted terms in the ligand field analysis---such as the $p_{\sigma}-p_{\pi}$ hopping---and multiplet effects are absent in the one-hole basis functions, the overall distribution of the hybridized states is qualitatively consistent with the data and is well captured by our \gls*{ED} calculations.

\section{Discussion}


Our detailed \gls*{RIXS} measurements and \gls*{ED} analysis put \La438{} in the mixed charge-transfer/Mott-Hubbard parameter regime of the \gls*{ZSA} scheme. Although the larger $\Delta$ in \La438{} suppresses the hybridization of the \gls*{TM} $3d$ and oxygen $2p$ orbitals, the fact that $\Delta$ is only about 2~eV larger, along with the enhanced $p-d$ hopping, guarantees that the oxygen orbitals retain a significant role in determining the physical properties of these low valence nickelates.  This also results in a substantial superexchange interaction that is half that of the cuprates \cite{Lin2021strong}. In this regard, low valence nickelates are a good analog of cuprate superconductors. However, the larger $\Delta$ produces a small, but possibly significant Ni $|d^8\rangle$ weight in the ground state.  


Since multi-orbital physics has been discussed extensively in nickelates, we compute the involvement of different Ni $d$ orbitals in the ground state. According to our \gls*{ED} calculations, the majority orbital character retains $d_{x^2-y^2}$ symmetry---less than 8\% of the holes occupy the $d_{3z^2-r^2}$ orbitals in \La438{} (see Appendix~\ref{wavefunction}). Hund's physics therefore has only a small role in determining the properties of this material, and the Ni states are predominantly of low-spin character, consistent with prior experiments \cite{Zhang2017large, Lin2021strong, Rossi2020orbital}. This is partly due to the strong \gls*{TM}-O hybridization, along with the square-planar environment of Ni, which gives rise to a large crystal field splitting in the $e_g$ states. Since both the $d_{x^2-y^2}$ and $d_{3z^2-r^2}$ orbitals hybridize with the O $2p_{\sigma}$ orbitals, they can mix despite their different symmetries. If $\Delta$ increases further, the $d_{3z^2-r^2}$ orbitals become more involved, as shown in Appendix~\ref{delta_dep}.

The mixed charge-transfer/Mott-Hubbard character of the low valence nickelates reported here suggests that minimal theoretical models must explicitly include both Ni and O states alongside strong correlations. In this regime, the properties of many-body methods such as \gls*{DMFT} can become dependent on the choice of basis functions to represent the correlations \cite{Karp2021dependence}. Problems with so-called ``double counting'' of $U$ are also more difficult to avoid in this case \cite{Dang2014covalency, Haule2015exact}. Both of these challenges likely contribute to the differences between different studies thus far \cite{Hu2019two,Nomura2019formation, Sakakibara2019model, Sawatzky2019superconductivity, Adhikary2020orbital, Bixia2020synthesis, Botana2020similarities, choi2020fluctuation, Hepting2020electronic, Jiang2020critical, Lechermann2020late, Lechermann2020multiorbital, Liu2020electronic, Petocchi2020normal,
Karp2020many, Kapeghian2020electronic, Nomura2020magnetic, Wang2020hunds, Werner2020nickelate, Wu2020robust, Zhang2020effective, Zhang2020self, Zhang2020type, Zeng2020phase, Been2021electronic, Goodge2021doping, Lang2021strongly, kang2021optical, Plienbumrung2021interplay, Wan2021exchange, Higashi2021core}. By combining O-resonant spectroscopy with an \gls*{ED} treatment of the relevant $\Delta$, $U$, and hopping parameters, we can confidently report that \La438{} is described by $U\sim\Delta$ with both values of order 6~eV. While our results position \La438{} in the mixed \gls*{ZSA} regime, other factors such as rare-earth orbitals, electron-phonon coupling, and La$_2$O$_2$ layers might contribute to the detailed properties of these materials \cite{Hepting2020electronic, Nomura2019formation, LaBollita2021electronic}. However, addressing these more subtle factors would expand the basis size well beyond what can be handled by current state-of-the-art \gls*{ED} calculations and require more approximate means of handling correlations, which is contrary to the goal of providing a rigorous minimal description of these materials.

\section{Conclusions}

In summary, we combine \gls*{RIXS} measurements at the O $K$-edge and \gls*{ED} calculations to compare the low valence nickelate \La438{} to a prototypical cuprate \LSCO{} with a similar nominal electron filling. Our work is unique in directly measuring the energy of the Ni-O hybridized states and interpreting them while treating $\Delta$, $U$, and hoppings on the same footing. The results reveal that \La438{} has a larger $\Delta$, smaller $U$, and increased $p-d$ hopping, which gives it a mixed charge-transfer/Mott-Hubbard electronic character. Despite its larger $\Delta$, the nickelate retains a sizable superexchange through its strong hopping. Moreover, for the parameters we have determined, Hund's physics is less relevant than might be expected (i.e., despite the admixture of $d^8$), consistent with the square-planar environment of Ni. Given that \La438{} and NdNiO$_2$ have similar magnetic exchange interactions \cite{Lin2021strong, Lu2021magnetic}, and since magnetic exchange is very sensitive to the $\Delta/U$ ratio, we suggest that this mixed charge-transfer/Mott-Hubbard picture likely applies to the entire $R_{n+1}$Ni$_{n}$O$_{2n+2}$ series. This idea is also supported by first-principles calculations, which find that these materials are indeed rather similar, provided they are compared at the same effective doping \cite{Karp2020comparative, LaBollita2021electronic}. Overall, we conclude that both the Coulomb interactions and charge-transfer processes need to be considered when interpreting the properties of these nickelates. Realistic models of low valence nickelates must therefore include both Ni and O states at a minimum, and we suggest this as a basis for conceptualizing these fascinating materials.

The supporting data for the plots in this article are openly available from the Zenodo database \cite{repo}.

\begin{acknowledgments}
We thank Kenji Ishii for helpful discussions. Work at Brookhaven National Laboratory (RIXS measurement and interpretation, and cuprate sample synthesis) was supported by the U.S. Department of Energy (DOE), Office of Science, Office of Basic Energy Sciences. Work at Argonne was supported by the U.S.\ DOE, Office of Science, Basic Energy Sciences, Materials Science and Engineering Division (nickelate sample synthesis and first-principles calculations). A.B.\ acknowledges support from NSF Grant No.~DMR 2045826. S.J.\ acknowledges support from the U.S. DOE, Office of Science, Office of Basic Energy Sciences, under Award No.~DE-SC0022311. X.H.\ was supported by the Gordon and Betty Moore Foundation's EPiQS Initiative through Grant No.~GBMF9074. This research used resources at the SIX beamline of the National Synchrotron Light Source II, a U.S.\ DOE Office of Science User Facility operated for the DOE Office of Science by Brookhaven National Laboratory under Contract No.~DE-SC0012704. This research used resources of the Advanced Photon Source, a U.S. DOE Office of Science User Facility at Argonne National Laboratory, and is based on research supported by the U.S. DOE Office of Science-Basic Energy Sciences, under Contract No.~DE-AC02-06CH11357.
\end{acknowledgments}

\appendix

\begin{table*}[htb]
\caption{Full list of parameters used for the \gls*{ED} calculations including the full set of \gls*{TM} $3d$ and oxygen $2p$ orbital energies. We set $\epsilon_{d_{x^2-y^2}}=0$, which means that, within hole notation, we need to set  $\epsilon_{p_{\sigma}}=\Delta$ \cite{supp}. Note that $V_{pd\sigma}$, $V_{pd\pi}$, $V_{pp\sigma}$, and $V_{pp\pi}$ are amplitudes of Slater-Koster parameters, and we fix $V_{pd\pi}=-V_{pd\sigma}/2$ and $V_{pp\pi}=-V_{pp\sigma}/4$. Correspondingly, $t_{p_{\sigma}p_{\sigma}}=(V_{pp\sigma}-V_{pp\pi})/2$ and $t_{p_{\sigma}d_{x^2-y^2}}=\sqrt{3}V_{pd\sigma}/2$. Here, $\eta$ is the ratio of the out-of-plane TM-O distance over the in-plane one, which controls the hopping amplitudes involving the apical oxygen. We assume $V_{pp\sigma}$ scales with $d^{-2}$, where $d$ is the O-O bond length, and $V_{pd\sigma}$ scales with $d^{-3.5}$, where $d$ is the TM-O bond length \cite{Eskes1990cluster}. Here, $F^0_{dd}$, $F^2_{dd}$, and $F^4_{dd}$ are Slater integrals for the \gls*{TM} $3d$ orbitals, and $F^0_{pp}$ and $F^2_{pp}$ are for oxygen $2p$. Correspondingly, $U=F^0_{dd}+4/49\times(F^2_{dd}+F^4_{dd})$, and the Hund's coupling is $J_H=(F^2_{dd}+F^4_{dd})/14$. Note that $U_{dp}$ is the inter-site Coulomb interaction between \gls*{TM} $3d$ and oxygen $2p$ holes, and $U_q$ is the core-hole potential for oxygen. Here, $\Delta$ and $U$ have an estimated error bar of about 1~eV \cite{supp}. All parameters, with the exception of $\eta$, are in units of eV. }
\begin{ruledtabular}
\begin{tabular}{cccccccccccccccccccc}
Cluster & $\epsilon_{d_{x^2-y^2}}$ & $\epsilon_{d_{3z^2-r^2}}$ & $\epsilon_{d_{xy}}$ & $\epsilon_{d_{xz/yz}}$ & $\epsilon_{p_{\sigma}}$ & $\epsilon_{p_{\pi}/p_z}$ & $\epsilon_{p^a_{\pi}}$ & $\epsilon_{p^a_z}$ & $V_{pd\sigma}$ & $V_{pp\sigma}$ & $\eta$ & $F^0_{dd}$ & $F^2_{dd}$ & $F^4_{dd}$ & $F^0_{pp}$ & $F^2_{pp}$ & $U_{dp}$ & $U_q$\\ 
\hline
Cu$_2$O$_{11}$  & 0 & 0.95 & 0.7 & 0.9 & 3.4 & 4 & 4 & 3.4 & 1.35 & 1 & 1.3 & 7.86 & 8.61 & 5.38 & 3.3 & 5 & 1 & 6 \\
Ni$_2$O$_7$ & 0 & 0.2 & 0.1 & 0.3 & 5.6 & 6.1 & - & - & 1.57 & 0.6 & - & 5.58 & 6.89 & 4.31 & 3.3 & 5 & 1 & 6 \\
\end{tabular}
\end{ruledtabular}
\label{table:allparams}
\end{table*}

\section{\label{synthesis}Sample synthesis and characterization}

The \La438{} single crystal used in this study was prepared as described in Refs.~\cite{Zhang2016stacked, Zhang2017large} and was the same piece as was used in Ref.~\cite{Lin2021strong}. The parent Ruddlesden-Popper \La438{} was prepared using the high-pressure optical floating zone method. During growth, oxygen gas was maintained at 20~bar pressure with a flow rate of 0.1~l/min. To improve homogeneity, the feed and seed rods were counter-rotated at 30~r.p.m., and rods were advanced at 4~mm/h over the 30~hour growth time. Sample reduction was performed by heating small crystals cleaved from the parent \La438{} boule in a flowing 4\% H$_2$/Ar gas mixture at 350$^{\circ}$C for five days. 

The \LSCO{} film reported here was synthesized using the atomic-layer-by-layer molecular beam epitaxy (ALL-MBE) technique \cite{Bozovic2001atomic}, in a similar way to samples used for prior \gls*{RIXS} studies \cite{Dean2012spin, Dean2013persistence, Meyers2017doping}. Our ALL-MBE system contains sixteen metal sources (thermal-effusion or Knudsen cells) and a source of pure (distilled) ozone. The growth is monitored in real time by means of reflection high-energy electron diffraction (RHEED), which provides information about the time evolution of the film surface morphology and crystal structure \cite{Bozovic1995analysis}. The growth kinetics is controlled using pneumatic linear-motion actuators that shutter the atomic sources. The \LSCO{} film studied here is heavily overdoped, $x = 0.35$. It was grown on a $10\times10$~mm$^2$ single-crystal LaSrAlO$_4$ substrate polished with the [001] axis perpendicular to the surface, with a miscut of less than $0.10^{\circ}$.  The LaSrAlO$_4$ substrate lattice constants are $a_0 = b_0 = 3.755$~\AA{}, $c_0 = 12.56$~\AA{}. The \LSCO{} film is 40 unit cells (530~\AA{}) thick, and it is pseudomorphic with the LaSrAlO$_4$ substrate; thus, it is under a small 0.5\% compressive strain \cite{Butko2009madelung}. Since LaSrAlO$_4$ has no pre-peak, it does not contribute to the measured signal. The substrate temperature during the growth (measured by a two-color pyrometer) was kept at $T = 611^\circ$C, and the ozone partial pressure was fixed at $p = 3\times10^{-5}$~Torr. The growth rate was about 0.03~\AA{}/sec. After the deposition, the film was annealed for 3.5~hours at $T = 611^\circ$C in an ozone partial pressure $p = 1\times10^{-4}$~Torr, and then cooled down slowly at the same pressure.

\begin{figure*}[tb]
\includegraphics{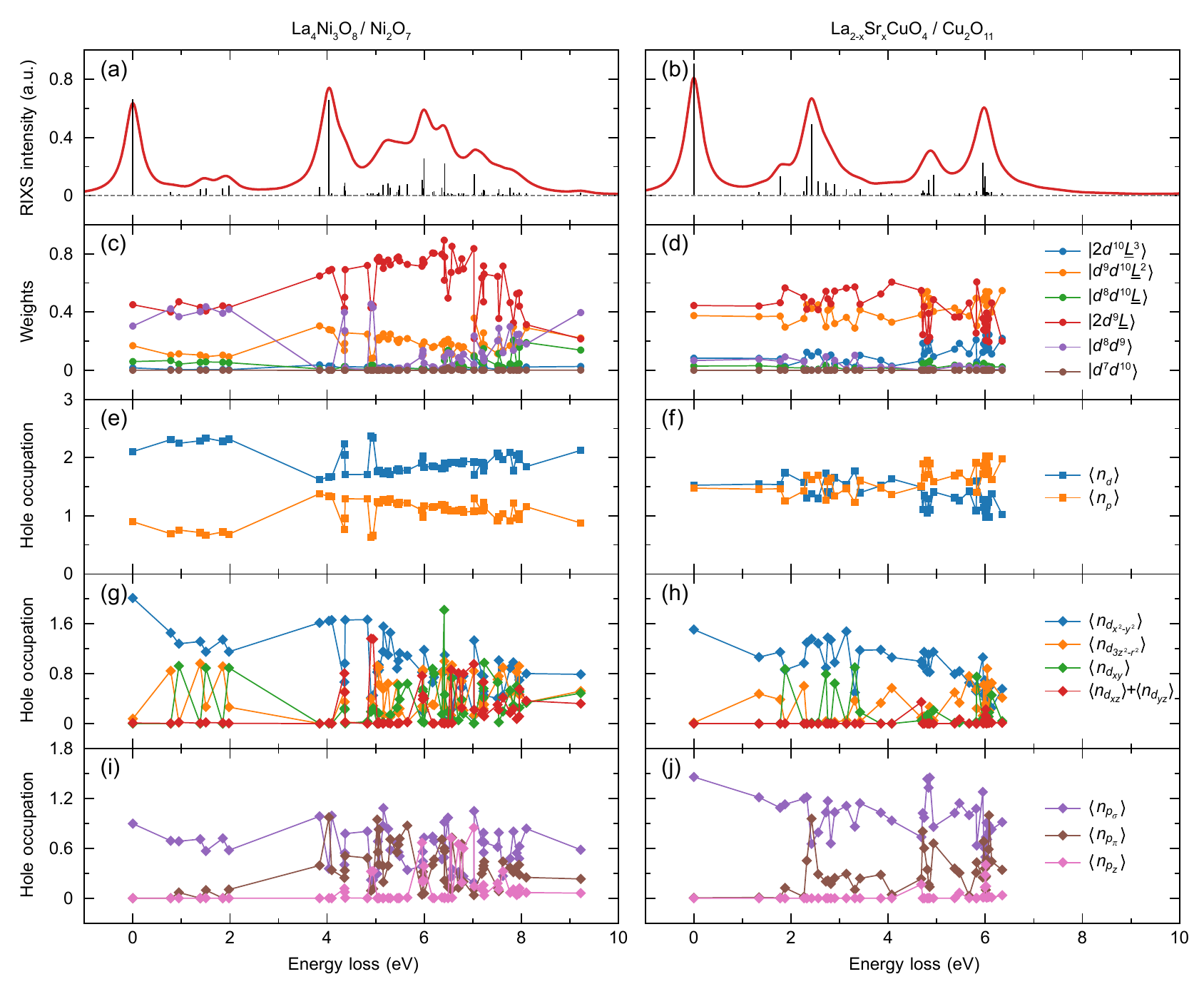}
\caption{Wave function analysis of Ni$_2$O$_7$ and Cu$_2$O$_{11}$ clusters with three holes. (a,b) Calculated unbroadened \gls*{RIXS} intensity (black vertical lines) and broadened \gls*{RIXS} spectra (red curves) with $\theta=60^\circ$ and $\sigma$ polarization. Note that in this geometry, the contribution from $p_z$ orbitals is minimized. (c,d) Weights of different configurations of the ground and excited states. (e,f) Hole occupations of \gls*{TM} $3d$ and O $2p$ orbitals. (g,h) Hole occupations for each type of $3d$ orbital. (i,j) Hole occupations for each type of $2p$ orbital. Only the excited states with unbroadened \gls*{RIXS} intensity stronger than 1\% of the maximum value are shown.}
\label{fig:wavefun}
\end{figure*}

\section{\label{notation}Notation of transition metal and oxygen orbitals}

For isolated atoms, the \gls*{TM} $3d$ and oxygen $2p$ orbitals are fully localized. In a compound, however, the hybridization among atoms will mix these orbitals, and the wave functions of the eigenstate orbitals are composed of different atomic-character Wannier orbitals. 

In the clusters used in this study, shown in Fig.~\ref{fig:EDcalc}(a) of the main text, the \gls*{TM} sites are all equivalent and include $d_{x^2-y^2}$, $d_{3z^2-r^2}$, $d_{xy}$ and $d_{xz/yz}$ orbitals. Oxygen atoms include $p_x$, $p_y$, $p_z$ orbitals, and they exist in inequivalent planar and apical locations; thus, we use $p$ and $p^a$ to distinguish them. The planar oxygen orbitals can be represented using $p_{\sigma}$ and $p_{\pi}$ orbitals that are parallel and perpendicular to the \gls*{TM}-O bonds, respectively, and that are composed of different $p_x$, $p_y$ and $p_z$ orbitals depending on the site. To further distinguish between the in-plane and out-of-plane $p_{\pi}$-type orbitals, we refer to the in-plane ones as $p_{\pi}$ and the out-of-plane ones as just $p_z$. For the apical oxygens, two sets of $p^a_{\pi}$ orbitals are equivalent, and $p^a_{\sigma}$ orbitals are labeled simply as $p^a_z$.

\section{\label{EDcalcs} Cluster exact diagonalization calculations}

For this work, we compute the \gls*{RIXS} cross section using the cluster \gls*{ED} approach. This accurately treats interactions and cross-section effects and is widely used to treat localized direct \gls*{RIXS} processes in correlated oxides \cite{Ament2011resonant, Okada2003theory, Okada2002copper, Okada2004effects, Monney2013determin, Monney2016probing, Johnston2016electron, Plienbumrung2021character, Chen2013doping}. This makes it appropriate for simulating features \textit{A} and \textit{B}, although it fails to capture x-ray fluorescence. The full Hamiltonian is composed of several terms 
\begin{equation}
    \mathcal{H}=\hat{E}_{d}+\hat{E}_{p}+\hat{\Delta}+\hat{U}_{dd}+\hat{U}_{pp}+\hat{U}_{dp}+\hat{U}_{q}+\hat{T}_{pp}+\hat{T}_{dp}
\end{equation}
where $\hat{E}_{d}$ and $\hat{E}_{p}$ are the on-site energies of \gls*{TM} $3d$ and oxygen $2p$ orbitals, respectively, and $\hat{\Delta}$ is the charge-transfer energy. The point-charge \gls*{CEF} splitting is included in these terms. Here, $\hat{U}_{dd}$ and $\hat{U}_{pp}$ describe the on-site Coulomb interactions for the \gls*{TM} and oxygen, respectively. All the Coulomb and exchange integrals are included explicitly. Note that $\hat{U}_{dp}$ is the inter-site Coulomb interaction between \gls*{TM} and oxygen. For the intermediate states, an additional term, $\hat{U}_{q}$, is included to account for the core-hole potential. Here, $\hat{T}_{pp}$ and $\hat{T}_{dp}$ refer to O-O $p-p$ hopping and \gls*{TM}-O $p-d$ hopping, respectively. All the hopping integrals are evaluated from the Slater-Koster parameters. The hopping terms involving apical oxygens are renormalized according to the bond lengths for the Cu$_2$O$_{11}$ cluster \cite{Eskes1990cluster}. The hopping phases are explicitly considered in the Hamiltonian. For simplicity, spin-orbit coupling is not considered. The full set of parameters used for calculating the spectra presented in the main text are listed in Table \ref{table:allparams}.

Regarding the Hilbert space, we include all the spin-resolved \gls*{TM} $3d$ and oxygen $2p$ orbitals and perform \gls*{ED} in the two-up, one-down spin sector for the doped clusters (three holes). For the undoped clusters (two holes), all the spin configurations are included. The Hamiltonian is constructed in the hole language. In this case, $\Delta$ is the energy difference between the $d_{x^2-y^2}$ and $p_{\sigma}$ orbitals regardless of the Coulomb interactions. In addition, by definition, $U$ equals the intra-orbital on-site Coulomb interaction for a $d^9$ system \cite{ZSA1985, ZSA1987}. The resulting eigenstates for the initial, intermediate, and final states from \gls*{ED} are used to calculate the \gls*{XAS} and \gls*{RIXS} spectra using the Kramers-Heisenberg formula in the dipole approximation with the experimental geometry explicitly considered. The inverse core-hole lifetime is fixed to 0.25~eV according to \gls*{XAS} data, and the final-state energy loss spectra are broadened using a Lorentzian function with $\sigma=0.2$ to account for the band width that is not captured by the small cluster. We sum up the spectra from clusters rotated by 90$^\circ$ around the $z$ axis to restore the $C_4$ symmetry. These calculations are compared with the experimental intensity after a simple self-absorption correction has been applied to the data based on the incident and emitted x-ray angles with respect to the surface \cite{Miao2017high}.

\section{\label{wavefunction}Wavefunction analysis of ED calculations}

To reveal the nature of the \gls*{RIXS} spectral features, we show in Fig.~\ref{fig:wavefun} the weights of different configurations of the excited states as well as the hole occupations. As discussed in the main text, feature A in \LSCO{} is dominated by the rearrangement of ligand orbitals, while for \La438{}, it involves a significant portion of charge transfer from Ni to O. In another aspect, feature B, which is mostly $dd$-excitations, is clearly identified in \La438{} since it is well separated from feature A. In \LSCO{}, it is weak and obscured by feature A.

\section{\label{ligand_analysis}Multiplet ligand field analysis}

\begin{table}
\caption{Energy-level splitting of the one-hole basis functions. The O $p$ orbitals (on-site energies $\epsilon_{p_{\sigma}}$, $\epsilon_{p_{\pi}}$, $\epsilon_{p_{z}}$) hybridize with each other to form ligand orbitals (energies $\epsilon_{L}$), which further hybridize with TM $d$ orbitals (energies $\epsilon_{d}$). Here, $V$ are the hopping integrals, and $T_{pp}=V_{pp\sigma}-V_{pp\pi}$. Note that the anti-bonding molecular 
$p_\pi$ ligand orbital that has an energy of $\epsilon_{p_{\pi}}-T_{pp}$ (not listed) will not hybridize with the TM $d$ states.}
\begin{ruledtabular}
\begin{tabular}{cccc}
Symmetry & $\epsilon_d$ & $\epsilon_L$ & $V$ \\ 
\hline
$x^2-y^2$  & $\epsilon_{d_{x^2-y^2}}$ & $\epsilon_{p_{\sigma}}-T_{pp}$ & $\sqrt{3}|V_{pd\sigma}|$ \\
$3z^2-r^2$  & $\epsilon_{d_{3z^2-r^2}}$ & $\epsilon_{p_{\sigma}}+T_{pp}$ & $|V_{pd\sigma}|$ \\
$xy$  & $\epsilon_{d_{xy}}$ & $\epsilon_{p_{\pi}}+T_{pp}$ & $2|V_{pd\pi}|$ \\
$xz/yz$  & $\epsilon_{d_{xz/yz}}$ & $\epsilon_{p_{z}}$ & $\sqrt{2}|V_{pd\pi}|$ \\
\end{tabular}
\end{ruledtabular}
\label{table:ligand}
\end{table}

Here, we provide a qualitative explanation of the multiplet ligand field analysis in Fig.~\ref{fig:THdep}(c). The analysis is the same as that derived more formally in Ref.~\cite{Eskes1990cluster}. For simplicity, we consider the hybridization of one hole in a NiO$_4$ cluster. When comparing to \La438{}, the sample's doping level is accounted for by offsetting the energy of these states when plotting the different hybridized orbital energies in Fig.~\ref{fig:THdep}(c). We start by considering the orbitals in the absence of  $p_{\sigma}-p_{\pi}$ hopping, such that the four $p_{\sigma}$ orbitals in a NiO$_4$ plaquette will hybridize with each other to form bonding and anti-bonding molecular orbitals. The non-bonding orbitals are not included \cite{Eskes1990cluster, Okada2002copper}. The same thing happens for $p_{\pi}$ orbitals, while $p_z$ orbitals remain non-bonding if the weak $p_z-p_z$ hopping is ignored. The bonding $p_\sigma$ molecular orbital has the same symmetry as the nickel $d_{3z^2-r^2}$ orbital, so they will hybridize with each other through $p_{\sigma}-d_{3z^2-r^2}$ hopping. The same applies for anti-bonding $p_\sigma$ molecular and $d_{x^2-y^2}$ orbitals, bonding $p_\pi$ and $d_{xy}$ orbitals, and the non-bonding $p_z$ and $d_{xz/yz}$ orbitals. The energy levels and hopping integrals are listed in Table~\ref{table:ligand}. Note that the anti-bonding $p_\pi$ orbital will not hybridize with nickel $d$ due to phase cancellation.

\section{\label{delta_dep}$\Delta$ dependence of RIXS spectra and hole occupation}

\begin{figure*}
\includegraphics{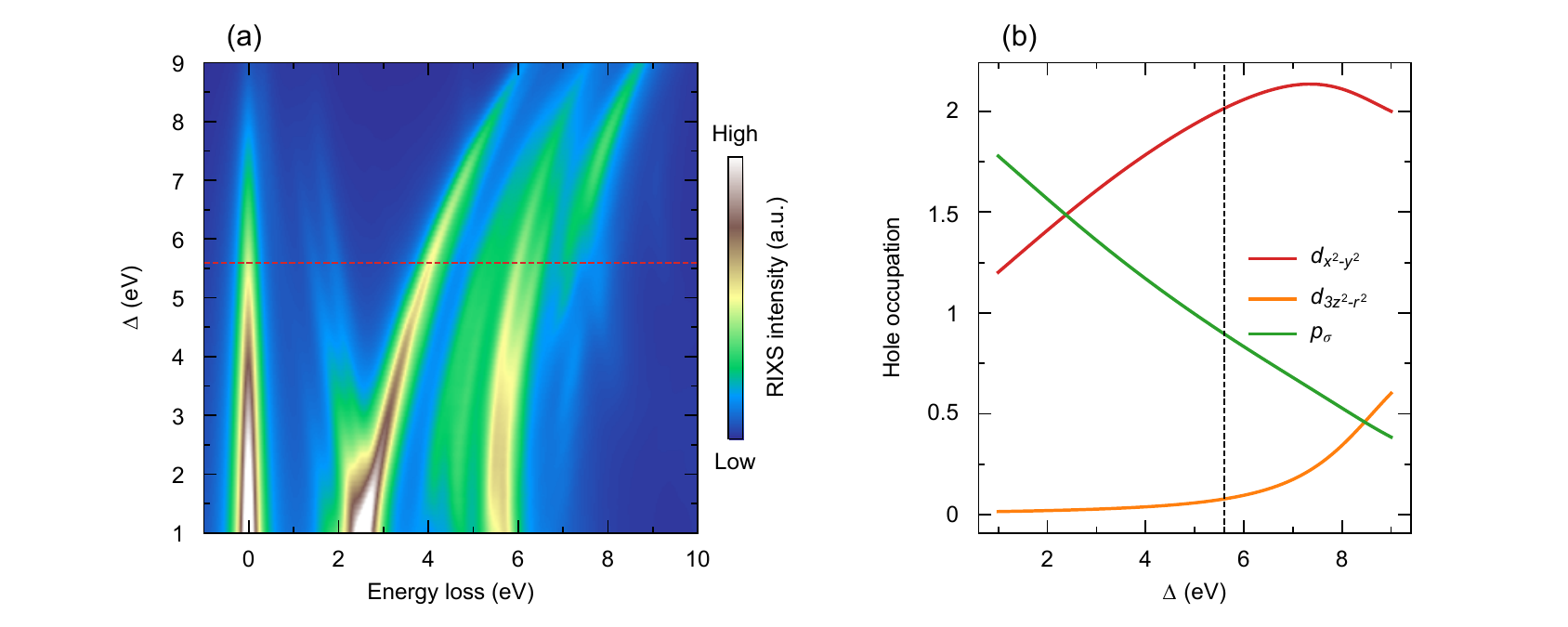}
\caption{Ni$_2$O$_7$ cluster calculations with different $\Delta$. Other parameters are fixed according to Table~\ref{table:allparams}. (a) Intensity map presenting the calculated $\Delta$-dependent \gls*{RIXS} spectra with $\theta=60^\circ$ and $\sigma$ polarization. (b) Orbital occupation of holes with different $\Delta$ in a doped cluster (three holes) with $U=6.5$~eV.}
\label{fig:Ddep}
\end{figure*}

Here, we present additional Ni$_2$O$_7$ cluster \gls*{ED} results to show how \gls*{RIXS} spectra and hole occupations evolve with $\Delta$. When $\Delta$ is small, only feature A can be resolved with an onset below 2~eV, resembling the cuprate data [see Fig.~\ref{fig:Ddep}(a)]. Concomitantly, the holes mainly occupy the oxygen sites [see Fig.~\ref{fig:Ddep}(b)]. With increasing $\Delta$, feature A shifts to higher energy, and feature B starts to emerge out of feature A. This also increases the number of holes transferred to the nickel sites. The red and black dashed lines indicate the $\Delta$ for \La438{} where feature A and feature B are well separated, but a significant portion of the holes remain on the oxygen sites. As $\Delta$ increases further, the $d_{x^2-y^2}$ hole concentration first increases, and then the $d_{3z^2-r^2}$ orbitals come into play, leading to multi-orbital physics.

\bibliography{refs}
\end{document}


\title{Supplemental Material: Role of oxygen states in low valence nickelate La$_4$Ni$_3$O$_8$}

\date{\today}

\maketitle

\renewcommand{\thefigure}{S\arabic{figure}}
\renewcommand{\thetable}{S\arabic{table}}

This document provides details on the definitions of $U$ and $\Delta$, electron versus hole notation, the choice of model parameters, the parameter dependence of the spectra, and the angular dependence of the spectra.

\section{Definitions of $U$ and $\Delta$}
Here we provide further details on the definition of the Coulomb and charge transfer energies. Since we need to accurately account for the many body effects, $U$ and $\Delta$ are defined in terms of specific electronic configurations in the atomic limit. $U$ is the energy required for a $d_i^nd_j^n\rightarrow d^{n-1}_id_j^{n+1}$ transition and $\Delta$ is defined in terms of a $d_i^n\rightarrow d^{n+1}_i\underline{L}$ transition, where $i$ and $j$ label \acrfull*{TM} sites and $\underline{L}$ denotes an oxygen ligand hole. In the present case, the reference occupation is $n=9$ electrons (or one hole). These definitions are the same as that of the widely used \gls*{ZSA} scheme \cite{ZSA1985}. We start by writing the Hamiltonian in hole language, which is the notation used throughout the main text:
\begin{equation}
H = \sum_{\sigma} \epsilon_d d_{\sigma}^\dagger d_{\sigma}^{\phantom\dagger} +
    \sum_{\delta,\sigma} \epsilon_p p_{\delta,\sigma}^\dagger p_{\delta,\sigma}^{\phantom\dagger} +
    U n^d_{\uparrow} n^d_{\downarrow} +
    \sum_{\delta} U_{pp} n^p_{\delta,\uparrow} n^p_{\delta,\downarrow}. \label{eq:H}
\end{equation}
For simplicity, this and the following equations involve only the $d_{x^2-y^2}$ and O $2p$ orbitals. The same analysis applies when the whole $d$ shell is included. In hole language, the relevant energies are
\begin{eqnarray}
    E(d^{0}p^0) &=& 0\\
    E(d^{1}p^0) &=& \epsilon_d \\
    E(d^{2}p^0) &=& 2 \epsilon_d + U \\
    E(d^{0}p^1) &=& \epsilon_p .
\end{eqnarray}
Coulomb repulsion follows the expected definition
\begin{equation}
U = E(d^{0}p^0) + E(d^{2}p^0) - 2 E(d^{1}p^0) = 2 \epsilon_d + U - 2 \epsilon_d = U.
\end{equation}
The charge transfer energy is 
\begin{equation}
\Delta = E(d^{0}p^1) - E(d^{1}p^0) = \epsilon_p - \epsilon_d
\end{equation}
consistent with the main text. Most literature on cuprates and the low valence nickelates use this hole notation.

\section{Electron versus hole notation}

How does Eq.~\ref{eq:H} change when we convert to electron language? In this case, we will use tilde notation to label terms in electron language. Applying the expected transformations to the operators, $H\rightarrow \tilde{H}$, $d_{\sigma}^{\phantom\dagger} \rightarrow \tilde{d}_{\sigma}^{\dagger}$, $n_{\sigma}^d \rightarrow (1-\tilde{n}_{\sigma}^d)$, etc.\ we obtain the Hamiltonian in electron language
\begin{equation}
\tilde{H} = \sum_{\sigma} \epsilon_d \tilde{d}_{\sigma}^{\phantom\dagger} \tilde{d}_{\sigma}^{\dagger} +
    \sum_{\delta,\sigma} \epsilon_p \tilde{p}_{\delta,\sigma}^{\phantom\dagger} \tilde{p}_{\delta,\sigma}^{\dagger} +
    U (1-\tilde{n}^d_{\uparrow})(1-\tilde{n}^d_{\downarrow}) +
    \sum_{\delta} U_{pp} (1-\tilde{n}^p_{\delta,\uparrow}) (1-\tilde{n}^p_{\delta,\downarrow}).
\end{equation}
We transform the first two terms via commutation rules to give
\begin{equation}
\tilde{H} = -\sum_{\sigma} \epsilon_d \tilde{d}_{\sigma}^{\dagger} \tilde{d}_{\sigma}^{\phantom\dagger} -
    \sum_{\delta,\sigma} \epsilon_p \tilde{p}_{\delta,\sigma}^{\dagger} \tilde{p}_{\delta,\sigma}^{\phantom\dagger} +
    U (1-\tilde{n}^d_{\uparrow})(1-\tilde{n}^d_{\downarrow}) +
    \sum_{\delta} U_{pp} (1-\tilde{n}^p_{\delta,\uparrow}) (1-\tilde{n}^p_{\delta,\downarrow}).
\end{equation}
Expanding the last two terms gives
\begin{equation}
\tilde{H} =  -\sum_{\sigma} (\epsilon_d+U) \tilde{d}_{\sigma}^{\dagger} \tilde{d}_{\sigma}^{\phantom\dagger} -
    \sum_{\delta,\sigma}  (\epsilon_p+U_{pp}) \tilde{p}_{\delta,\sigma}^{\dagger} \tilde{p}_{\delta,\sigma}^{\phantom\dagger} +
    U \tilde{n}^d_{\uparrow} \tilde{n}^d_{\downarrow} +
    \sum_{\delta} U_{pp} \tilde{n}^p_{\delta,\uparrow} \tilde{n}^p_{\delta,\downarrow},
\end{equation}
where some constant terms have been omitted. By comparison to Eq.~\ref{eq:H} we can associate $\tilde{\epsilon}_d=-(\epsilon_d+U)$, $\tilde{\epsilon}_p=-(\epsilon_p+U_{pp})$,
$\tilde{U}=U$, and $\tilde{U}_{pp}=U_{pp}$. This follows the expectation that a particle-hole transformation leaves the Coulomb interactions unchanged but shifts the on-site energies by the value of the local Hubbard repulsion, reflecting a change in the vacuum state. Our result in hole language, $\Delta=\epsilon_p-\epsilon_d$, can then be recast as $\Delta=-(\tilde{\epsilon}_p+U_{pp})+(\tilde{\epsilon}_d+U)$ in electron language, which is the same as $\tilde{\Delta}=E(d^{10}\underline{L})-E(d^9)$. This transformation shows that when comparing with first principles calculations, care needs to be taken about how the correlations and many body effects are handled. To do so, one must take into account possible double counting errors \cite{Dang2014covalency, Haule2015exact}, which will effectively change the meanings and values of $\epsilon_d$ and $\epsilon_p$.

\section{The choice of model parameters in ED calculations}

Here we provide the details about how the parameters are determined/chosen in the \gls*{ED} calculations. Although there are many parameters in the original Hamiltonian, most of them are constrained by the physical considerations described below so that they can be effectively fixed with our quoted accuracy of 1~eV. We categorize the parameters in four parts.

\begin{enumerate}

    \item \textbf{Coulomb interactions.} Since we work in the hole language, the on-site Coulomb repulsion for O is not crucial since double hole occupation on the O sites is unlikely. Therefore, $F^0_{pp}$ and $F^2_{pp}$ do not influence our conclusions and are fixed to standard Hartree-Fock values throughout the calculations \cite{Wang2019EDRIXS}. The inter-site Coulomb interaction $U_{dp}$ is also fixed since the distance between Ni and O atoms means that it is expected to be much smaller than the Ni on-site Coulomb interactions and thus plays a negligible role \cite{Okada1997intersite}. The ratio between $F^4_{dd}$ and $F^2_{dd}$ is known to be approximately 5/8, independent of solid-state screening \cite{deGroot19902p, deGroot2008core}. The Hund’s exchange in \gls*{TM} materials is known empirically to vary between 0.4 and 1.2~eV \cite{deGroot19902p, deGroot2008core}. Hence, it is safe to fix it with an estimated uncertainty much smaller than 1~eV. By doing so, the only tunable parameter with the largest uncertainty is the Ni intra-orbital Coulomb repulsion $U$. Once $U$ is chosen, $F^0_{dd}$, as well as the inter-orbital Coulomb interactions, is uniquely determined through our procedure.
    
    \item \textbf{Hopping integrals.} Making use of the Slater-Koster scheme, all the hopping integrals can be derived from two parameters, $V_{pd\sigma}$ and $V_{pp\sigma}$. The other two parameters are set as $V_{pd\pi}=-V_{pd\sigma}/2$ and $V_{pp\pi}=-V_{pp\sigma}/4$ based on the known scaling of the hopping for \gls*{TM}-O octahedra of this type \cite{Mizuno1998Electronic}. This fixes these two parameters with an accuracy better than our estimated error bar. An additional parameter, $\eta$, is used for the Cu$_2$O$_{11}$ cluster, since it has apical O bonds, and this is fixed to the standard value \cite{Eskes1990cluster}.
    
    \item \textbf{Point charge crystal field splitting.} The \gls*{CEF} splitting is primarily inferred from the low energy $dd$ excitations and has less of an effect on the charge transfer ones. It has two contributions. The major one is the $p-p$ and $p-d$ hoppings and the minor one is the point charge \gls*{CEF}. Therefore, once the hopping integrals are chosen, the point charge \gls*{CEF} parameters can be confidently determined to match the $dd$ excitations seen at both the O $K$ and Ni $L$ edges with an uncertainty much smaller than 1~eV.
    
    \item \textbf{Other parameters.} The core-hole potential $U_q$ only modifies the peak intensities but not the peak positions since it is not involved in the ground states and the excited states. So it is fixed to a standard value for the O $K$-edge of 6~eV \cite{Okada2006copper}. The charge transfer energy $\Delta$ is the most important quantity we want to extract, so we treat this as fully tunable.
    
\end{enumerate}

Points 1-4 leave $U$, $\Delta$, $V_{pd\sigma}$, and $V_{pp\sigma}$, as the only free parameters. These have distinct effects on the \gls*{RIXS} spectra, allowing them to be determined by comparison with the \gls*{RIXS} spectra.  The width of feature A along the energy loss axis is primarily determined by $V_{pp\sigma}$ while its position is a correlated combination of $U$, $\Delta$, and $V_{pd\sigma}$, which is further constrained by the superexchange interaction $J$. We also need to consider the relative intensities of features within the manifold of feature A as well as the ratio between features A and B along with their angular dependence. With these constraints, we search the parameter space to match our RIXS data [Fig.~S1], and the results are presented in Fig.~3 and Fig.~S2. The primary factor limiting the accuracy is a range of different $\Delta$ and $U$ values that produce reasonable agreement with the data and which gives us our estimated error bar of 1~eV. We test our approach by measuring and applying the same analysis to a cuprate. As explained in the main text, the fact that we obtain values in good accord with the literature validates the accuracy of our approach within our quoted error bar.  

\begin{figure}[ht!]
\includegraphics{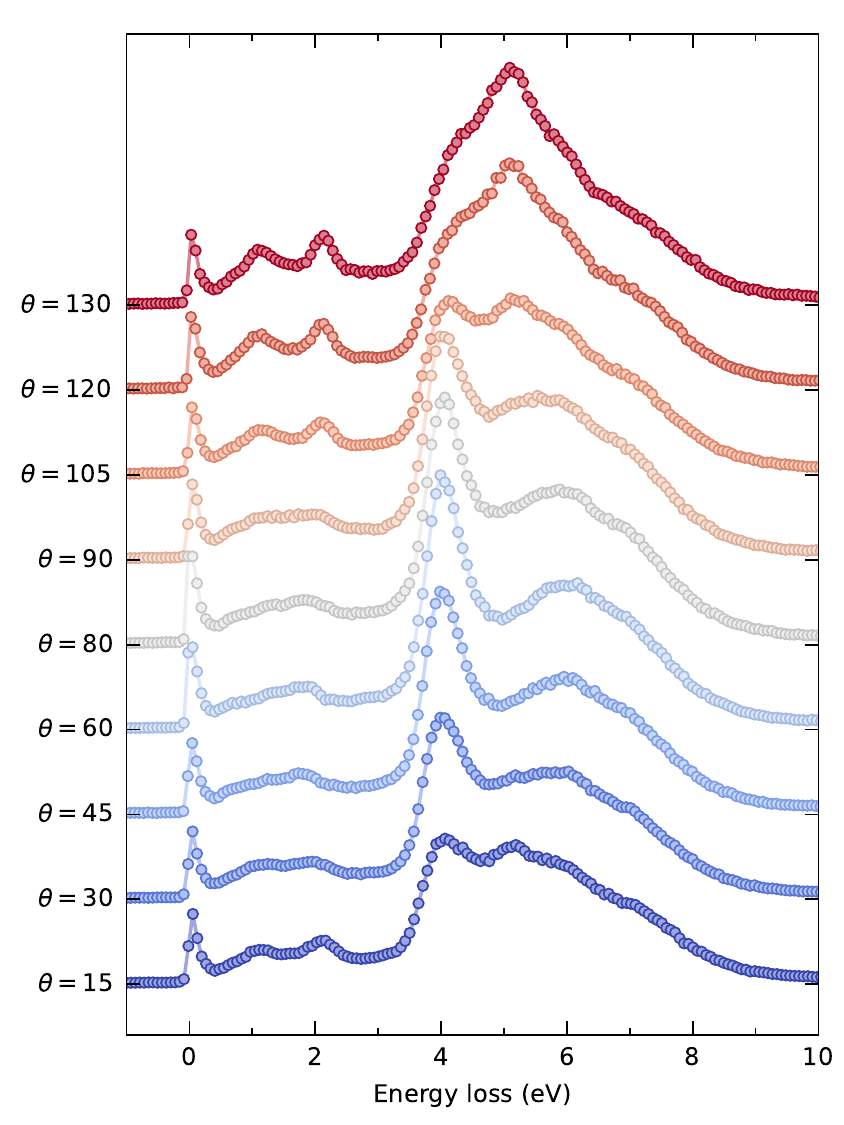}
\caption{Self-absorption corrected \gls*{RIXS} spectra of \La438{} collected at 40~K. The same set of data produce Fig.~4(a) in the main text.}
\label{fig:THdep_raw}
\end{figure}

\begin{figure}[ht!]
\includegraphics{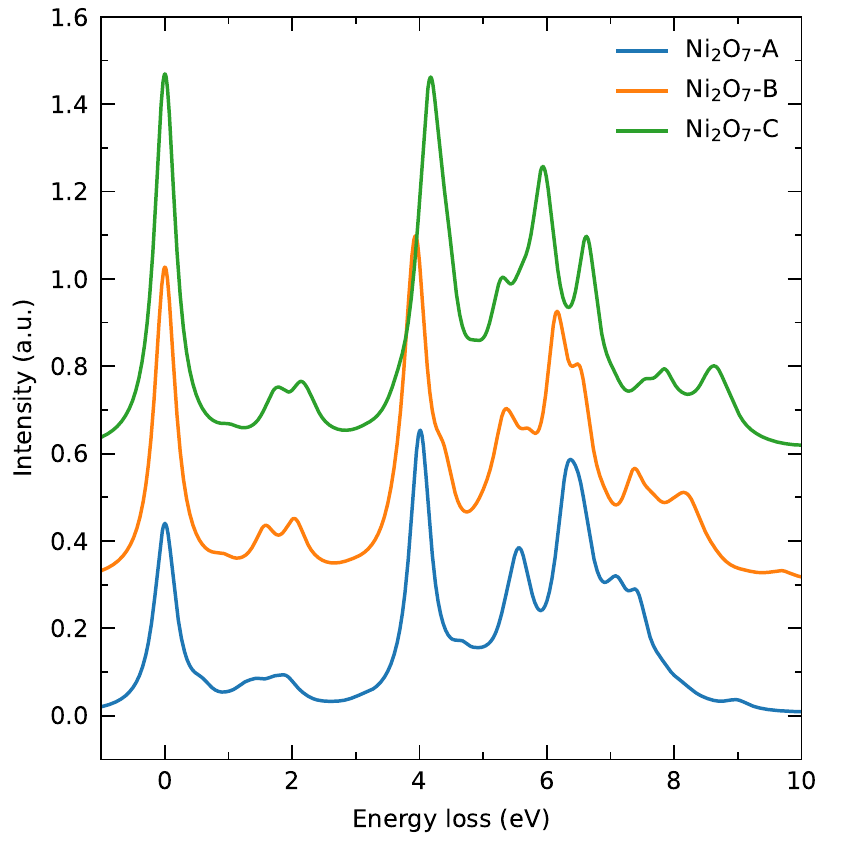}
\caption{Calculated \gls*{RIXS} spectra of Ni$_2$O$_7$ cluster using the parameters listed in Table~\ref{table:moreparams}. The calculations are performed with $\theta=60^\circ$ and $\sigma$ polarization. The spectra are shifted along the $y$ axis for clarity.}
\label{fig:moreEDcalc}
\end{figure}

\begin{table*}[ht!]
\caption{List of parameters used for the Ni$_2$O$_7$ \gls*{ED} calculations in Fig.~\ref{fig:moreEDcalc}. All parameters are listed in units of eV.}
\begin{ruledtabular}
\begin{tabular}{ccccccccccccccccc}
Label & $\epsilon_{d_{x^2-y^2}}$ & $\epsilon_{d_{3z^2-r^2}}$ & $\epsilon_{d_{xy}}$ & $\epsilon_{d_{xz/yz}}$ & $\epsilon_{p_{\sigma}}$ & $\epsilon_{p_{\pi}/p_z}$ & $V_{pd\sigma}$ & $V_{pp\sigma}$ & $F^0_{dd}$ & $F^2_{dd}$ & $F^4_{dd}$ & $F^0_{pp}$ & $F^2_{pp}$ & $U_{dp}$ & $U_q$ & $J_{\textrm{calc}}$\\ 
\hline
A & 0 & 0.2 & 0.3 & 0.5 & 5.8 & 5.8 & 1.55 & 0.6 & 4.58 & 6.89 & 4.31 & 3.3 & 5 & 1 & 6 & 0.083\\
B & 0 & 0.2 & 0 & 0.2 & 6 & 6.5 & 1.63 & 0.7 & 6.58 & 6.89 & 4.31 & 3.3 & 5 & 1 & 6 & 0.085\\
C & 0 & 0.2 & 0 & 0.2 & 5.9 & 6.9 & 1.65 & 0.7 & 7.58 & 6.89 & 4.31 & 3.3 & 5 & 1 & 6 & 0.085\\
\end{tabular}
\end{ruledtabular}
\label{table:moreparams}
\end{table*}

\section{Parameter dependence of the spectra}

To test the accuracy of our derived parameters, we investigated the variation of the spectra when changing the input parameters. Table~\ref{table:moreparams} and Fig.~\ref{fig:moreEDcalc} show the range of different $\Delta = \epsilon_{p_{\sigma}}$ values that produce a satisfactory fit of the spectra. We conclude that $\Delta\approx 6$~eV within an error bar of about 1~eV, such that \La438{} can be robustly classified as a mixed charge-transfer/ Mott-Hubbard system with $\Delta \sim U$. For \LSCO{}, we were able to further constrain our values using literature O $K$-edge \gls*{RIXS} measurements of the undoped sample \cite{Bisogni2012Bimagnon}, so the values for this material are constrained to $\lesssim~1$~eV. This approach is currently impractical for nickelates as efforts to electron dope \La438{} have so far not been successful, and infinite-layer nickelate films are inhomogeneous due to spatial variability in the oxygen reduction \cite{Lu2021magnetic, Rossi2020orbital, Fu2019core, Hepting2020electronic, Gu2020single, Goodge2021doping, Chen2021electronic}.  

\section{Angular dependence of O $K$-edge spectra}

In this section we explain the angular dependence of the spectra. Figure 4 in the main text was taken at a fixed scattering angle of $2\Theta=150^{\circ}$ while varying the incident angle $\theta$. In this way, the angle between the outgoing beam and the sample surface is $\theta_\text{out} = \theta +30^\circ$. The incident x-ray polarization was $\sigma$ and therefore parallel to the sample $b$-axis at all $\theta$, whereas the emitted x-ray polarization was not analyzed such that both $\pi$ and $\sigma$ polarizations are detected. To understand the overall angular dependence we calculate the transition probability out of the $p_\sigma$, $p_\pi$, and $p_z$ orbitals at different $\theta$, which captures the main transitions, even though this leaves out some detailed structure that is present in the full calculations in Fig.~4. Excitations involving $p_\pi$ are maximized when $\theta_\text{out}=90^\circ$ (equivalent to $\theta=60^\circ$) such that all emitted photons are polarized in the $ab$ plane of the sample. Excitations involving $p_z$ are largest as $\theta_\text{out}$ gets close to $180^\circ$ such that the $\pi$ polarized emitted photons are along the sample $c$-axis. 

\begin{figure}[ht!]
\includegraphics{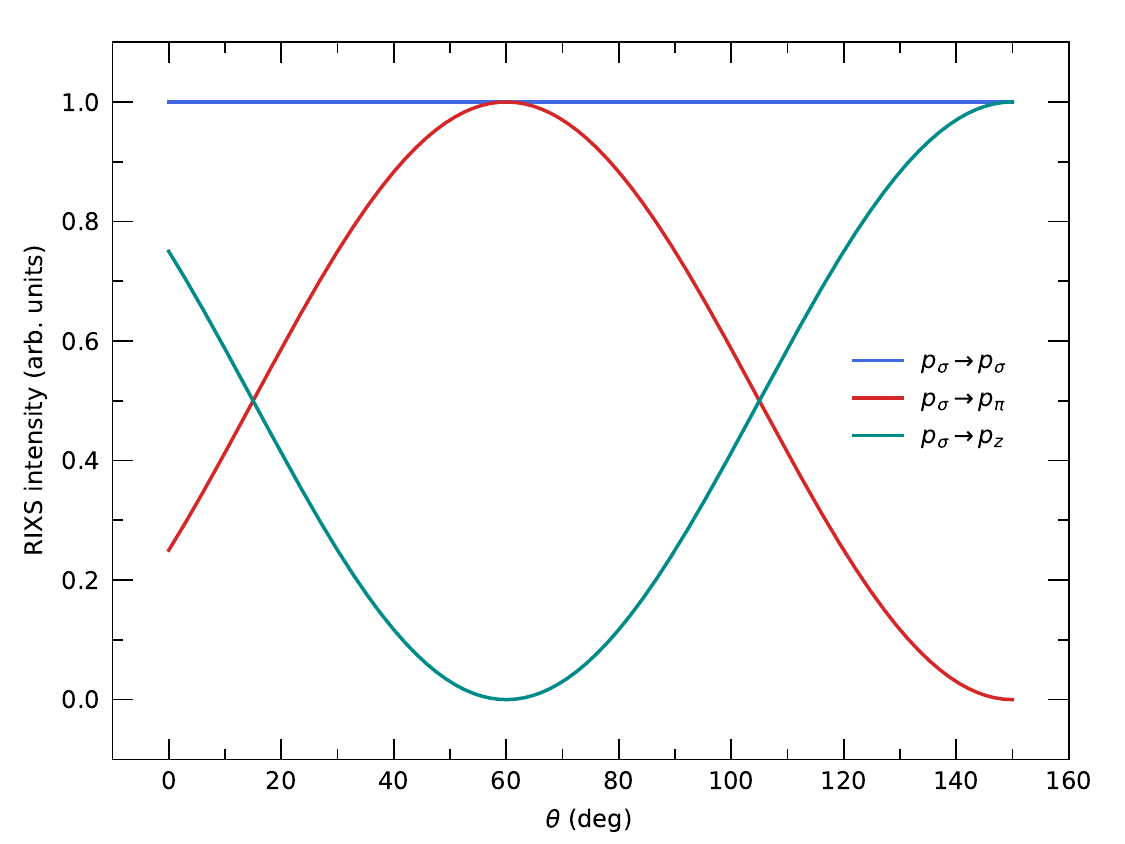}
\caption{Calculated \gls*{RIXS} intensities for transitions between different combinations of $p_\sigma$, $p_\pi$, and $p_z$ as a function of incident angle $\theta$ with $\sigma$ polarized incident x-rays.}
\label{fig:angular_dep}
\end{figure}

\clearpage
\bibliography{refs}